\documentclass[12pt,preprint2]{aastex}
\shorttitle{Determination of the orbit of HD~45350~b.}
\shortauthors{Endl M., Cochran W.D., Wittenmyer R.A., Hatzes A.P.}
\begin{document}

\title{Determination Of The Orbit Of The Planetary Companion To The Metal Rich Star HD 45350
\footnote{
Based on observations obtained with the Harlan J. Smith Telescope and the 
Hobby-Eberly Telescope, which is a joint project of the 
University of Texas at Austin, the Pennsylvania State University, Stanford University, 
Ludwig-Maximilians-Universit\"at M\"unchen, and Georg-August-Universit\"at G\"ottingen.
}}
\author{Michael Endl, William D. Cochran, Robert A. Wittenmyer} 
\affil{McDonald Observatory, The University of Texas at Austin, Austin, TX 78712}
\email{mike@astro.as.utexas.edu, wdc@astro.as.utexas.edu, robw@astro.as.utexas.edu}
\and
\author{Artie P. Hatzes}
\affil{Th\"uringer Landessternwarte, D - 07778 Tautenburg, Germany}
\email{artie@tls-tautenburg.de}

\begin{abstract}

We present the precise radial velocity (RV) data for the metal-rich star HD~45350 collected with
the Harlan J. Smith (HJS) 2.7~m telescope and the Hobby-Eberly Telescope (HET) at McDonald Observatory. 
This star was noticed by us as a candidate for having a giant planetary companion in a highly eccentric orbit, 
but the lack of data close to periastron left the amplitude and thus mass of the planet poorly constrained.      
Marcy et al.~(2005) announced the presence of the planet based on their Keck/HIRES data,
but those authors also cautioned that the remaining uncertainties in the orbital solution might be large due 
to insufficient data near periastron passage. In order to close this phase gap
we exploited the flexible queue scheduled observing mode of the HET
to obtain intensive coverage of the most recent periastron passage of the planet. 
In combination with the long term data from the HJS 2.7~m telescope we determine a Keplerian
orbital solution for this system with a period of $962$~days, an eccentricity of $e=0.76$ and a velocity 
semi-amplitude $K$ of $57.4~{\rm m \, s}^{-1}$. The planet has a minimum mass of 
$m\sin i = 1.82 \pm 0.14~{\rm M}_{\rm Jup}$ and an orbital semi-major axis of
$a = 1.92\pm0.07$~AU.
\end{abstract}


\section{Introduction}

The majority of the extrasolar planets discovered by Doppler surveys (e.g. Marcy et al.~2005)
with periods beyond the tidal circularization limit of a few days reside in eccentric orbits. 
Circular planetary orbits with periods of several years appear to be rather exceptional. 
Orbits with certain 
combinations of high eccentricity and longitudes of periastron can pose severe difficulties 
to the observer to characterize the true shape of the radial velocity (RV) orbit, because the 
maximum change in velocity near the periastron passage exhibits itself as a very sharp and 
narrow peak despite a long orbital period. All the information on the RV amplitude and thus on the minimum 
mass of the companion is contained in this part of the RV orbit. 
It is necessary to sample this
orbital phase as densely as possible to determine the best orbital solution.
Knowledge of the most precise orbital parameters for extrasolar planets can play a crucial role in 
the discovery and study of multi-planetary systems (e.g. McArthur et al.~2004, Rivera et al.~2005).    

The HD~45350 planetary system represents such a case, where the high eccentricity and the 
periastron angle leads to the steep rise and fall of the RV orbit near periastron having  
a short duration of only a few weeks, despite an overall orbital period of more than $2.5$ years. 
The star was added to the target list of our Doppler planet search survey
at the Harlan J. Smith (HJS) 2.7~m telescope (e.g. Cochran et al. 1997, Hatzes et al. 2003, Endl et al. 2004)
as a promising planet host star, based on its high metallicity, as suggested by Laughlin~(2000). 
Soon we detected a parabolic RV acceleration indicating a possible planetary 
companion. A preliminary orbital solution yielded $P\approx973$~days and 
$m\sin i = 1.5~{\rm M}_{\rm Jup}$. However, 
the remaining uncertainties were large due to the insufficient sampling of the RV orbit
close to periastron and through the RV maximum. 
Marcy et al.~(2005) announced the presence of the planet based on their Keck/HIRES data and also 
noted that the remaining uncertainties of the orbital parameters (especially of the eccentricity) 
likely exceed the formal errors, because of the lack of data near periastron.

In this work we present the orbital solution and the $m\sin i$ value for the planet orbiting 
HD~45350 based on observations at McDonald Observatory. 
We were able to determine this solution by closely monitoring the most recent periastron passage and 
following the orbit through the RV maximum. 
Most of the data which sample this orbital phase were obtained with the HET, where we used the 
queue scheduled observing mode to achieve a dense sampling of the RV orbit, independent of our observing run 
schedule at the HJS 2.7~m telescope.

\section{Parameters of the host star HD~45350}

HD~45350 (= HIP 30860) is a $V=7.9$ magnitude G-type star at a distance of 
$48.9$~pc according to the Hipparcos parallax of $20.43$~mas (ESA 1997).    
Laughlin~(2000) determined a high metallicity of [Fe/H]$=0.19$ and proposed the 
star as a promising planet host based on its high metal content. 
Allende Prieto \& Lambert (1999) give a stellar mass of $0.98~{\rm M}_{\odot}$, while
Marcy et al.~(2005) adopt a slightly higher value of $1.02\pm0.1~{\rm M}_{\odot}$ based
on its high metallicity. For the same reason we also adopt the higher value for 
the primary mass.

Our HJS 2.7~m spectra also include the Ca~II~H\&K lines, which can be used as a measure for the
chromospheric emission and serve as proxies for the overall magnetic activity level. The 
McDonald S-index for HD~45350 is $0.159\pm0.034$, consistent with an old quiescent star where
activity modulations of RV data are negligible. This generally low activity level
for HD~45350 was also found by Marcy et al.~(2005).    

\section{Observations of HD~45350 at McDonald Observatory}

RV monitoring of HD~45350 using the HJS 2.7~m telescope at McDonald observatory
started in 2000 November. We have collected a total of $45$ measurements as of 2006 January 
with this telescope. The precise RV results are obtained with the
standard I$_2$-cell technique (e.g. Endl, K\"urster, \& Els~2000). 
The left hand column of table~\ref{vels} lists our HJS 2.7~m telescope RV measurements for HD~45350. 
The data have a total rms scatter of $31.5~{\rm m \, s}^{-1}$ and a mean formal internal error of 
$8.9~{\rm m \, s}^{-1}$. 

At the end of 2003 we noticed a turnaround in the overall RV trend of HD~45350 which apparently was caused by 
a highly eccentric orbit, where we had just missed the periastron because the star was unobservable.
A preliminary orbit predicted the next periastron passage at the end of 2005. We therefore added   
HD~45350 to our observing queue at the HET in order to follow the RV orbit through this periastron
passage and RV maximum. The right hand columns of table~\ref{vels} contains the $28$ velocity 
measurements from the HET. 
Instrument configuration and data reduction procedures for the HET data are described in detail 
in Cochran et al.~(2004).

\section{The orbital solution of the system}  

The program $Gaussfit$ (Jefferys et al.~1988) 
has proven to be a powerful tool to determine orbital solutions by combining several different RV data 
sets (e.g. Hatzes et al.~2003). We used $Gaussfit$ to find a Keplerian orbital solution to the combination of 
the HJS 2.7~m telescope and HET data, including the arbitrary zero-points ($\gamma$ velocities) of both data sets as 
fit parameters in the model. We determined an orbital solution with the following parameters:
$P=962.1\pm4.4$~days, $K=57.4\pm1.8~{\rm m \, s}^{-1}$, $e=0.764\pm0.01$, 
${\rm T}_{\rm per}=2451827.4\pm9.4$~JD, and $\omega=338.7\pm4.1$~degrees. The fit yields a $\chi^{2}$ of 
$89.3$ and for $66$ degrees of freedom a $\chi^{2}_{\rm red}$ of $1.35$. Figure 1 displays our RV data
along with our Keplerian orbital solution. Figure 2 shows an expanded view of the RV maximum near periastron
passage. 

We derive a mass function for this system of $f(m)=5.0636\pm0.2032 \times 10^{-9}~{\rm M}_{\odot}$.
Assuming a mass for the host star of $1.02\pm0.1~{\rm M}_{\odot}$ we derive a minimum mass for the planet 
of $m\sin i = 1.82\pm0.14~{\rm M}_{\rm Jup}$ at an orbital semi-major axis of ${\rm a} = 1.92 \pm 0.069$~AU.
Due to the high eccentricity the planet has a periastron distance to the star of only $0.453$~AU and the
apastron is at $3.39$~AU. Table 3 summarizes the orbital parameters of the system. 

The total rms scatter around the orbit is $9.05~{\rm m \, s}^{-1}$, the residual rms scatter of the
HJS 2.7~m data alone is $10.3~{\rm m \, s}^{-1}$ and of the HET data $6.7~{\rm m \, s}^{-1}$. Both values are
slightly larger than the formal internal errors. There are several possible 
explanations for this: the star might be intrinsically variable, 
there are additional companions in this system, 
or the errors of some spectra are underestimated (or even a combination of all these effects). 
As mentioned before, HD~45350 shows no sign of an enhanced 
chromospheric activity, which makes the first explanation unlikely. In the next section we will discuss the 
possibility of additional companions. Since we have used all available spectra of HD~45350, taken at different 
airmasses and during varying seeing conditions at McDonald Observatory, and did not perform any   
pre-selection (e.g.  according to signal-to-noise ratio) of the data set, we think the 
third explanation is the most likely one.    
 
\subsection{Comparison to the Keck data}

A comparison of the orbit derived from the McDonald data with the published $30$ Keck/HIRES velocities from
Marcy et al.~(2005) is shown in Figure 3. The residual rms scatter of $5.5~{\rm m \, s}^{-1}$ is now 
slightly higher than the rms scatter around the orbit presented by Marcy et al. ($4.6~{\rm m \, s}^{-1}$),
but the Keck data are consistent with the orbit we have determined. 

We have also performed a combined orbital solution using both the two McDonald data sets and the Keck data. The
orbital parameters derived from using all three available data sets are also given in Tab. 3. 
This solution is consistent within the uncertainties of the orbit determined with the McDonald data alone.
 
\section{Searching for additional planets in this system}

Because the residual scatter around the orbit is slightly larger than expected we performed 
a period search to look for indications of additional companions in this system. The 
Lomb-Scargle periodogram (Lomb 1976 ; Scargle 1982) of the McDonald Observatory RV residuals is shown in Figure 4. 
The highest point in the power spectrum is at a period of $4.07$~days, but has an estimated false-alarm-probability of
$29\%$. Due to the lack of any statistically significant signal in the residuals we see no
evidence for a second (inner) planet in this system. The small periastron distance of the 
$\approx 1.8~{\rm M}_{\rm Jup}$ planet probably creates a dynamically unstable region for planets inside
$0.4$~AU. A dynamical study of this system will show at what semi-major axes planetary orbits interior to
HD~45350~b remain stable for a significant period of time.
 
In order to detect additional exterior planets, it will take many more observations of the full orbital cycle 
of the first planet to find modulations of the RV maxima in time. Again, a dynamical study should show at what
separation planetary orbits outside the first companion will remain stable. 
 
\section{Conclusion}

We determined the precise orbit of the planet around HD~45350. The peculiar shape of the RV orbit, caused
by the combination of high eccentricity and longitude of periastron, made it difficult to determine the
amplitude and thus the mass of the companion without good sampling of the RV maximum.
Marcy et al.~(2005) carefully noted that the remaining uncertainties of 
their orbital parameters likely exceed their formal errors, because of 
the lack of Keck data near periastron. Our results confirm their suspicion since the $m\sin i$ value of
the planet is almost twice the value they have obtained. 
We also performed a search for periodic signals in the residuals from the orbit to look for additional 
inner companions, but found no statistically significant periodicities. 

This result demonstrates again that the HET is a powerful tool to perform time critical observations of
extrasolar planetary systems. The observations of HD~45350 were similar to the case of
HD~37605 (Cochran et al.~2004), where a highly eccentric orbit made dense sampling of crucial orbital 
phases necessary.  

\acknowledgements
We thank the referee Geoff Marcy for his positive comments on the manuscript.
We are grateful to the McDonald Observatory Time-Allocation-Committee for generous allocation of 
observing time. This material is based upon work supported by
the National Aeronautics and Space Administration under Grants NNG04G141G and NNG05G107G.
The Hobby-Eberly Telescope (HET) is a joint project of the University of Texas at Austin, the 
Pennsylvania State University, Stanford University, Ludwig-Maximilians-Universit\"at M\"unchen, 
and Georg-August-Universit\"at G\"ottingen. 
The HET is named in honor of its principal benefactors, William P. Hobby and Robert E. Eberly.

\begin{deluxetable}{rrrrrr}
\tabletypesize{\tiny}
\tablecaption{
Differential radial velocities for HD~45350 from the HJS 2.7~m telescope 
and HET}
\tablewidth{0pt}
\tablehead{
& \colhead{HJS 2.7~m} & & & \colhead{HET}\\
\colhead{JD [2,400,000+]} & \colhead{dRV [${\rm m\,s}^{-1}$]} & 
\colhead{$\sigma$ [${\rm m\,s}^{-1}$]} & 
\colhead{JD [2,400,000+]} & \colhead{dRV [${\rm m\,s}^{-1}$]} &
\colhead{$\sigma$ [${\rm m\,s}^{-1}$]}  
\label{vels} }
\startdata
 51862.80903  &     55.2  &   9.54 & 53628.96521 &  -44.6  &   3.95\\ 
 51918.85994  &     16.9  &   8.48 & 53633.92327 &  -42.2  &   4.80\\
 51920.78122  &     30.0  &   8.52 & 53663.85563 &  -44.5  &   3.59\\
 51987.70230  &     10.0  &   8.53 & 53675.83402 &  -36.0  &   3.63\\
 52185.00333  &    -17.4  &   8.09 & 53679.81294 &  -36.1  &   3.63\\ 
 52249.88814  &    -14.2 &    8.59 & 53682.81009 &  -30.4  &   3.79\\
 52328.72557  &    -10.0  &   8.50 & 53683.79466 &  -24.9  &   3.64\\ 
 52331.64911  &     -9.1  &   8.80 & 53685.80867 &  -28.8 &    4.02\\
 52354.65723  &    -21.7  &   8.12 & 53701.76042 &  -18.3  &   4.43\\
 52386.61170  &    -25.9  &   9.38 & 53706.74234 &  -20.1  &   4.34\\
 52541.95920  &    -28.2  &   8.52 & 53708.76689 &  -20.9  &   3.90\\
 52576.92374  &    -19.8 &    9.05 & 53710.94995 &  -17.4  &   3.57\\
 52577.94127  &    -31.6  &   9.03 & 53712.96788 &  -17.5  &   4.98\\
 52597.94784  &    -17.8  &   7.94 & 53712.97523 &  -11.7  &   5.54\\
 52599.95755  &    -27.1 &    8.34 & 53719.92874 &  -19.6  &   4.44\\
 52600.96018  &    -27.9  &   7.65 & 53721.71659 &   -7.3  &   4.23\\
 52619.91330  &     -9.8  &   8.37 & 53723.68371 &  -13.4  &   4.73\\
 52658.84819  &     -2.5 &    7.99 & 53725.93705 &    1.6  &   4.03\\ 
 52661.63313  &     -1.9 &   10.14 & 53727.92978 &    6.9  &   4.18\\ 
 52688.80484  &    -33.2 &    9.74 & 53731.68362 &    6.6  &   3.93\\      
 52688.81984  &    -16.9 &    8.55 & 53734.66350 &   15.9  &   4.58\\ 
 52742.66242  &      6.6 &    9.03 & 53738.64973 &   34.4 &    4.31\\
 52742.67395  &      2.0  &   8.98 & 53749.85322 &   66.0  &   4.27\\ 
 52743.63340  &     11.4  &   8.21 & 53753.85074 &   67.3 &    4.45\\ 
 52743.64146  &      9.2  &   8.91 & 53754.62678 &   69.3 &    4.33\\
 52930.94794  &     19.1  &   8.54 & 53758.60945 &   66.2 &    4.76\\
 52959.93133  &      3.2  &   8.57 & 53765.81048 &   61.6 &    4.76\\ 
 53015.82992  &     11.2 &   14.81 & 53771.59327 &   45.5 &    5.62\\
 53035.79226  &      6.8  &   9.09 & & & \\
 53067.65125  &     -6.9  &   8.15 & & & \\
 53073.67958  &     -2.1  &   8.36 & & &\\
 53319.96635  &    -20.1  &   9.16 & & &\\
 53392.72016  &     -5.7  &   9.03 & & &\\
 53423.78226  &    -21.4  &   8.30 & & &\\
 53425.76844  &    -28.7 &    9.32 & & &\\
 53425.78190  &    -43.9  &  10.01 & & &\\
 53433.68634  &    -39.8  &   7.95 & & &\\
 53464.64883  &    -19.1  &  10.94 & & &\\
 53465.61972  &     -4.2  &   8.25 & & &\\
 53655.93450  &     10.3  &   9.54 & & &\\
 53691.96418  &     -1.4  &   8.45 & & & \\
 53745.81221  &     78.8  &   9.48 & & &\\
 53745.85423  &     79.4  &   9.04 & & &\\
 53746.82785  &     91.5  &   8.37 & & &\\
 53748.78906  &     66.8   &  9.34 & & &\\
\enddata
\end{deluxetable}

\clearpage

\begin{deluxetable}{rrcrcl}
\tabletypesize{\scriptsize}
\tablecaption{
The orbital parameters for the  HD~45350 planetary system}
\tablewidth{0pt}
\tablehead{
\colhead{parameter} & \colhead{value} & 
\colhead{uncertainty} & \colhead{value} & \colhead{uncertainty} & \colhead{unit}\\
\colhead{} & \colhead{HJS+HET} & \colhead{HJS+HET} & \colhead{HJS+HET+Keck} & \colhead{HJS+HET+Keck} & 
\label{orbit} }
\startdata
Period & 962.1 & 4.4 & 963.6& 3.4 & days \\
K & 57.4 & 1.8 & 58.0 & 1.7 & ${\rm m \, s}^{-1}$\\
e & 0.764 & 0.011 & 0.778 & 0.009 & \\
T$_{\rm per}$ & 2451827.4 & 9.4 &2451825.3  &7.1 & JD\\
$\omega$ & 338.7 & 4.1 & 343.4 & 2.3 & degrees\\
$f(m)$ & 5.0636 & 0.2032 & 4.8305 & 0.1847 & $\times 10^{-9}~{\rm M}_{\odot}$ \\
$m\sin i$ & $1.82$ & $0.14$ & $1.79$ & $0.14$ & ${\rm M}_{\rm Jup}$\\
$a$ & 1.92 & $0.069$ & 1.92 & 0.067 & AU \\ 
\enddata
\end{deluxetable}

\clearpage

\begin{figure} 
\includegraphics[angle=270,scale=0.3]{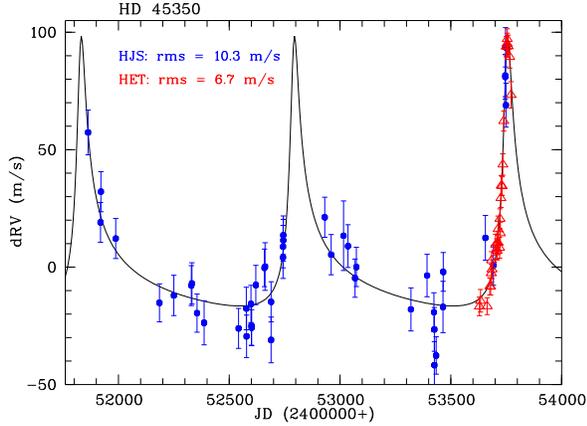}
\caption{RV measurements for HD~45350 (G5~V), from the HJS 2.7~m telescope (dots) and from 
the HET (triangles). The solid line displays our best-fit orbital solution (see table 3 for the
orbital parameters).
        }
\label{orbit_time}
\end{figure}

\begin{figure} 
\includegraphics[angle=270,scale=0.3]{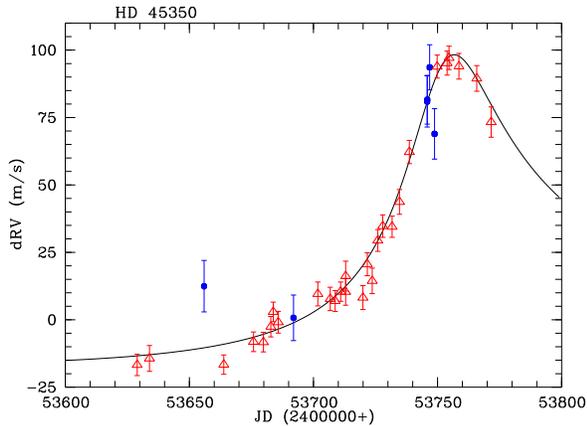}
\caption{
Expanded view of figure 1 showing the critical orbital phase near the periastron passage of the 
planet. Again, the dots represent HJS 2.7~m data and the triangles are the HET data. 
The stars changes its RV by $\approx 100~{\rm m \, s}^{-1}$ in about $61$ days despite an overall
orbital period of $962$~days.}
\label{orbit_time2}
\end{figure}

\begin{figure} 
\includegraphics[angle=270,scale=0.3]{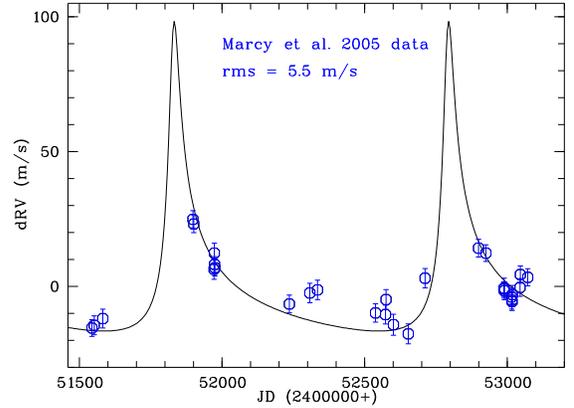}
\caption{
Comparison of the orbital solution based on the McDonald Observatory data (solid line) to the
published Keck/HIRES velocities from Marcy et al.~2005. 
        }
\label{orbit_time3}
\end{figure}

\begin{figure} 
\includegraphics[angle=270,scale=0.3]{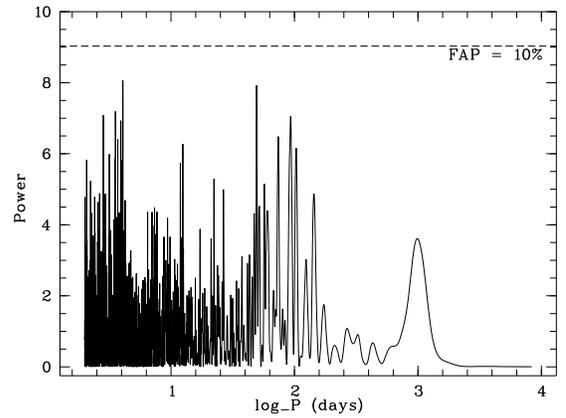}
\caption{
Lomb-Scargle periodogram of the RV residuals (HJS + HET) after subtraction of the best-fit 
orbital solution. The horizontal dashed line displays the power level corresponding to a false-alarm-probability
(FAP) of $10\%$. The highest peak in the power spectrum is around 4 days and has a FAP of $29\%$. 
        }
\label{periodo}
\end{figure}

\end{document}